\documentclass[12pt,a4paper]{article}
\usepackage{latexsym,amsmath,amssymb}
\date{}
\numberwithin{equation}{section}

\begin{document}
\title{Supersymmetric Canonical Commutation Relations \\}
\author{Florin Constantinescu\\ Fachbereich Mathematik \\ Johann Wolfgang Goethe-Universit\"at Frankfurt\\ Robert-Mayer-Strasse 10\\ D 60054
Frankfurt am Main, Germany }
\maketitle

\begin{abstract}
We discuss unitarily represented supersymmetric canonical commutation relations which are subsequently used to canonically quantize massive and massless chiral, antichiral and vector fields. The canonical quantization shows some new facets which do not appear in the non supersymmetric case. Our tool is the supersymmetric positivity generating the Hilbert-Krein structure of the $N=1$ superspace. 
\end{abstract}

\section{Introduction}

Supersymmetric fields are introduced either as multiplets of usual non-supersymmetric field components or by path integral methods \cite{WB}. The canonical quantization of supersymmetry doesn't seem to have been worked out. We proceed here with the first steps studying supersymmetric commutation relations for supersymmetric creation and annihilation operators and subsequently use them to quantize massive and massless free supersymmetric chiral, antichiral und vector fields. Our SUSY commutation relations are unitarily represented in symmetric Fock spaces constructed over supersymmetric test functions with vacuum being the supersymmetric function one. Our tool is the Hilbert-Krein structure introduced in \cite{C1,C2}. The multiplet components are usual functions of space-time. There is no difference between "bosonic" and "fermionic" components \cite{C2} of the supersymmetric test functions to which a variant of the van der Waerden calculus has to be applied. The Hilbert-Krein structure provides a frame for unitarity of the supersymmetric canonical commutation relations. Unitarity means that the commutation relations are represented in terms of operators acting in a positive definite (Hilbert) space. As expected the massless case is more interesting giving rise to some novelties. In the second preparatory section we fix the notations, shortly explain the approach in \cite{C1} and provide some further tools. In the third section we give the canonical quantization relations and their invariant Fock space representations. In the fourth section we quantize supersymmetric fields by canonical commutation relations. In section five we pass to the massless case and insist on the massless vector field. Some conclusions are drawn in the last section .\\ 
The main point of the paper is the study of rigorous unitarily represented, manifestly supersymmetric canonical commutation relations. They are used to canonically quantize free supersymmetric fields in positive definite Hilbert (Fock) spaces. There is no problem with non-propagating auxiliary field because they are eliminated by their algebraic equations of motion in the process of test function factorization conducting to the (on-shell) Hilbert-Krein structure. In principle we obtain equivalence with path integral methods but the massless vector case shows new interesting points. \\
Before ending the introduction we want to provide the reasons for writting this paper. The canonical quantization in quantum field theory is a method in which fields appear as operators in a (positive definite) Hilbert space (the Fock space) whose elements are the states (including the vacuum) of the system. In supersymmetry our canonical approach turns out to be unusual because states contain Grassmann variables and are no longer associated to real particles. But the formalism itself seems to work without problems. This could be of real interest for problems like unitary S-matrix theory which are outside reach of path integrals if we go beyond formal considerations. The interest on alternative methods is even more stringent if we want to test by methods with a higher degree of rigorousity remarkable achivements steeming from the "non-renormalization" theorems of supersymmetry. 

\section{Supersymmetric positivity}

We use the notations in supersymmetry in \cite{S}. The notations and formulas in non-supersymmetric quantum field theory are entirely borrowed from the treatise \cite {W}. \\
The most general supersymmetric function will be denoted by

\begin{gather}\nonumber 
X(z)=X(x,\theta ,\bar \theta )= \\ \nonumber
=f(x)+\theta \varphi (x) +\bar \theta \bar \chi (x) +\theta ^2m(x)+\bar \theta^2n(x)+ \\ 
+\theta \sigma^l\bar \theta v_l(x)+\theta^2\bar \theta \bar \lambda(x)+\bar \theta^2\theta \psi (x)+ \theta^2 \bar \theta^2d(x)
\end{gather}
where the coefficients are (complex) functions of $x$ of certain regularity (for instance elements of the Schwartz space $ S $). Using the covariant AND invariant \cite{S} p.45 derivatives:

\begin{gather}  
D_{\alpha }=\partial_{\alpha } +i\sigma_{\alpha \dot \alpha }^l\bar
\theta^{\dot \alpha }\partial _l \\ 
D^{\alpha }=\epsilon ^{\alpha \beta }D_{\beta }=-\partial ^{\alpha }+i\sigma^{l\alpha }_{\dot \alpha }\bar \theta^{\dot \alpha }\partial _l  \\
\bar D_{\dot \alpha }=-\bar {\partial }_{\dot \alpha } -i\theta ^{\alpha }\sigma _{\alpha \dot \alpha }^l\partial _l \\ 
\bar D^{\dot \alpha }=\epsilon ^{\dot \alpha \dot \beta }\bar D_{\dot
  \beta }=\bar {\partial }^{\dot \alpha }-i\theta ^{\alpha }\sigma
_{\alpha }^{l\dot \alpha }\partial _l   
\end{gather}
we define chiral, antichiral and transversal functions by the conditions
\[ \bar D^{\dot \alpha }X=0, \dot \alpha =1,2;  D^{ \alpha }X=0, \alpha =1,2; D^2 X=\bar D^2 X=0 \]
respectively. From now on in this section we restrict to the massive case. Let $p_0 =\sqrt{\bar p^2 +m^2 }, \bar p =(p_1 ,p_2 ,p_3 )$ and let \cite{W}

\begin{gather}\nonumber
\Delta_ + =\Delta_+ (m)= \\
=\frac{1}{(2\pi)^3}\int \frac{d^3 p }{2p_0 }e^{-ipx}=\frac{1}{(2\pi )^3}\int d^4 p \theta (p_0 )\delta (p^2 +m^2 )e^{-ipx}
\end{gather}
be the scalar field two point function. The difference

\begin{gather}
\Delta (x)= \Delta_+ (x)-\Delta_+ (-x)
\end{gather}
is the scalar field commutator. Define
 
\begin{gather}
K_0(z_1 ,z_2)=K_0(z_1 -z_2)=\delta^2 (\theta_1 -\theta_2)\delta^2 (\bar \theta_1 -\bar \theta_2 )\Delta_+ (x_1 -x_2)
\end{gather}
and

\begin{gather}
K_c (z_1 ,z_2)=P_c K_0 (z_1 -z_2 )\\
K_a (z_1 ,z_2)=P_a K_0 (z_1 -z_2) \\
K_+ (z_1 ,z_2 )=P_+ K_0 (z_1 -z_2 )\\
K_- (z_1 ,z_2 )=P_- K_0 (z_1 -z_2 )  \\
K_T (z_1 ,z_2)=-P_T K_0 (z_1 -z_2)
\end{gather}
where $P_i ,i=c,a,T,+,- $ are defined in \cite{WB} and act here on the first variable. 
They are 

\begin{gather}\nonumber
P_c =\frac{1}{16\square }\bar D^2 D^2 ,P_a =\frac{1}{16\square }D^2 \bar D^2 \\ \nonumber 
P_T =-\frac{1}{8\square }D^{\alpha }\bar D^2 D_{\alpha }=-\frac{1}{8\square }\bar
D_{\dot \alpha }D^2 \bar D^{\dot \alpha }  \\ \nonumber 
P_+ =\frac{D^2 }{4\sqrt {\square }}, P_- =\frac{\bar D^2 }{4\sqrt {\square }}
\end{gather}
For $i=c,a,T $ the operators $P_i $ are projections with $P_i P_j =0, i\neq j $ and $ P_c +P_a +P_T =1 $. Define invariant inner products for $i=c,a,T,+,- $ as:

\begin{gather}
(X_1 ,X_2 )_i =\int \bar X_1 ^T (z_1 ) K_i (z_1 ,z_2 )X_2 (z_2 )d^8 z_1 d^8 z_2
\end{gather}
and

\begin{gather}
(X_1 ,X_2 ) =\int \bar X_1 ^T (z_1 ) K (z_1 ,z_2 )X_2 (z_2 )d^8 z_1 d^8 z_2
\end{gather}
where in (2.15) $X$ is no longer a function but a vector with components $P_c X ,P_a X ,P_T X $ and $K$ the matrix kernel 

\begin{gather}
\begin{pmatrix} P_c & P_+ & 0 \\
  P_- & P_a& 0 \\ 0 & 0 & -P_T
\end{pmatrix}K_0
\end{gather}
They satisfy the complex conjugation property of inner products $(X_1 ,X_2 )=\overline {(X_2 ,X_1 )} $ where the bar means merely complex conjugation. In (2.15) $\bar X^T $ is the transpose conjugate of $ X $. It includes Grassmann besides complex conjugation. Note that the integration in (2.14), (2.15) includes Grassmann variables too (Berezin integrals).  
It was proved in \cite{C1}, section 2 and \cite{C2} that the inner products defined above with the exception of $ (.,.)_i ,i=+,- $ are positive definite. 
In order to obtain strict positive definiteness we must factorize the zero vectors of the kernel (2.16). This test function factorization is equivalent \cite{C1,C2}, by well-known duality between test functions and distributions, with imposing equations of motions on fields including the algebraic ones for the auxiliarx fields. This will be the reason for the fact that auxiliary non-propagating field do not cross the process of canonical quantization explained in Section 4. 
Note the minus sign in the definition of the kernel $K_T $ in (2.13) and in (2.16). In connection with (2.13) it is responsible for a genuine Krein structure of the space of supersymmetric functions which together with the induced Hilbert space was called in \cite{C1} the standard supersymmetric Hilbert-Krein structure. The absence of the minus sign in (2.13) would destroy  unitarity and as such would be a disaster. Indeed the diagonal projections in (2.16) would summ up to one and the form (2.15) would be indefinite as it was proved in \cite{C1}, section 2. Generally the positive definiteness over Grassmann algebra is a delicate issue. This makes the Hilbert-Krein structure above even more interesting. The reason for the positive definiteness of (2.14) and (2.15) is that after all Berezin integrals are done only positive biliner forms of bosonic and fermionic quantities separately are left \cite{C1}. They summ up to give the supersymmetric scalar product. The interesting consequence is that the Hilbert space of a multiplet is not isomorphic to the Hilbert space tensor product of the multiplet components, as we might be inclined to assume, but to the Hilbert space direct sum. Ones the intrinsec Hilbert space of supersymmetric functions has been found (it will play the role of the "one-particle" space), supersymmetric operators are treated as usual Hilbert space operators. This includes domain of definitions and hermiten adjoints. The only points which have to be mentioned is that the conjugation includes both complex and Grassmann and that the fermionic components of the superfunctions appearing as elements in the Hilbet space are commutative (cf.\cite{C2}). The last point produces some changes in the corresponding van der Waerden calculus and is the reason of good news \cite{C2} concerning Hilbert space aspects of supersymmetry; in particular supersymmetric operators. The odd part in the conjugated $\bar X $ of (2.1) changes sign \cite{C2}, (2.13) and as such the supersymmetric partial integration \cite{S}, section 8.2 is consistent with the definition of the adjoint operator. This is not the case in the usual SUSY formalism \cite{S}, section 8.2; in fact in the usual formalism even the covariant derivatives have no operatorial meaning being nasty, nonmathematical objects \cite{C2}, (2.21).

\section{Canonical commutation relations}

We pass now to the supersymmetric canonical commutation relations which we induce by using the above positive definite scalar products on test function superspace. The creation and annihilation operators appearing in this section act on superfunctions of the form (2.1) with regular coefficients (for instance in the Schwartz space S). We start with chiral commutation relations for supersymmetric creation and annihilation operators $a,a^*,b,b^* $. They are defined through their commutation relations

\begin{gather}\nonumber
[a(p_1 ,\theta_1 ,\bar \theta_1 ),a^* (p_2 ,\theta_2  ,\bar \theta_2 )]= \\  
=\frac{1}{16p_0 }\bar D^2 D^2 \delta^3 (p_1 -p_2 )\delta^2 (\theta_1 -\theta_2 )\delta^2 (\bar \theta_1 -\bar \theta_2 ) \\ 
[b(p_1 ,\theta_1 ,\bar \theta_1 ),b^* (p_2 ,\theta_2  ,\bar \theta_2 )]= \\ 
=\frac{1}{16p_0 } D^2 \bar D^2 \delta^3 (p_1 -p_2 )\delta^2 (\theta_1 -\theta_2 )\delta^2 (\bar \theta_1 -\bar \theta_2 )
\end{gather}
and

\begin{gather}\nonumber
[a(p_1 ,\theta_1 ,\bar \theta_1 ),b(p_2 ,\theta_2  ,\bar \theta_2 )]= \\
=\frac{m\lambda }{4p_0 }\bar D^2 \delta^3 (p_1 -p_2 )\delta^2 (\theta_1 -\theta_2 )\delta^2 (\bar \theta_1 -\bar \theta_2 )\\ 
[a^* (p_1 ,\theta_1 ,\bar \theta_1 ),b^* (p_2 ,\theta_2  ,\bar \theta_2 )]= \\
=\frac{m\lambda }{4p_0 } D^2 \delta^3 (p_1 -p_2 )\delta^2 (\theta_1 -\theta_2 )\delta^2 (\bar \theta_1 -\bar \theta_2 )
\end{gather}
supplemented by the trivial commutators
\[ [a(p_1 ,\theta_1 ,\bar \theta_1 ),a(p_2 ,\theta_2  ,\bar \theta_2 )]=[b(p_1 ,\theta_1 ,\bar \theta_1 ),b(p_2 ,\theta_2  ,\bar \theta_2 )]=0 \]
\[ [a^* (p_1 ,\theta_1 ,\bar \theta_1 ),a^* (p_2 ,\theta_2  ,\bar \theta_2 )]=[b^* (p_1 ,\theta_1 ,\bar \theta_1 ),b^* (p_2 ,\theta_2  ,\bar \theta_2 )]=0 \]
where in momentum space

\begin{gather}
D_{\alpha }=\frac{\partial }{\partial \theta^{\alpha }}-\sigma_{\alpha \dot \alpha }^l \bar \theta^{\dot \alpha }p_l \\ 
\bar D_{\dot \alpha }=-\frac{\partial }{\partial \bar \theta^{\dot \alpha }}+\theta^{\alpha }\sigma_{\alpha \dot \alpha }^l p_l 
\end{gather}
and $\lambda $ is an auxiliary non-negative constant which at a certain moment will be restricted to zero or one. For $\lambda =0$, in order to have a name, we call the commutation relations (and the chiral and antichiral field generated by them in the next section) "non-diagonal".
In the smeared form the non-trivial commutation relation (3.1)-(3.4) read

\begin{gather} 
[a(X),a^* (Y)]=(\bar X ,Y)_c  \\
[b(X),b^* (Y)]=(\bar X ,Y)_a  \\
[a(X),b(Y)]=\lambda (\bar X ,Y)_+  \\
[a^* (X),b^* (Y)]=\lambda (\bar X ,Y)_-
\end{gather}
where $ (.,.)_i ,i=c,a,+,- $ are sesquilinear forms introduced above and $X,Y$ test superfunctions. The operators $a,a^* ,b,b^* $ are unitary represented in symmetric Fock spaces of supersymmetric functions. It is constructed as usual over the Hilbert space of "one-particle" states. The vacuum is the supersymmetric function one. Certainly this vacuum has nothing to do with the Clifford vacuum used in superymmetry to represent the supersymmetric Poincare group \cite{WB}. The operators $a^* ,b^* $ are Hilbert space adjoints of $a,b$. \\
We introduce a third category of commutation relations (used in the vector case to follow)

\begin{gather}\nonumber
[c(p_1 ,\theta_1 ,\bar \theta_1 ),c^* (p_2 ,\theta_2  ,\bar \theta_2 )]= \\  
=\frac{1}{p_0 }(-\xi_T  P_T +\xi_c  P_c +\xi_a  P_a ) \delta^3 (p_1 -p_2 )\delta^2 (\theta_1 -\theta_2 )\delta^2 (\bar \theta_1 -\bar \theta_2 )
\end{gather}
supplemented by
\[[c(p_1 ,\theta_1 ,\bar \theta_1 ),c(p_2 ,\theta_2  ,\bar \theta_2 )]=0\]
where $P_i, i=c,a,T $ are projections introduced in Section 2 and $\xi_i ,i=T,c,a $ auxiliary constants. Sound choices will be $\xi_T =1, \xi_c =\xi_a =0$ and $\xi_T =\xi_c =\xi_a =1$ in which cases positivity is assured. In the particular case  $\xi_T = -1,\xi_c =\xi_a =1$ the projections sum up to one but this choice is at odd with positivity as already stated. The geometric locus of $\xi $ which renders (3.12) positive (more precisely non-negative) is the set $\xi_i \geq 0 $. It was shown in \cite{C1} section 3 that this ambiguity is related to families of gauges of the supersymmetric vector field. For the convenience of the reader this aspect will be reconsidered in Section 4 in the context of the gauge symmetry of canonical quantization. It also appears in the integral formalism (see \cite{WB} p.73). As in the case of $a,b,a^* ,b^* $ in (3.7)-(3.10) we can define the smeared version of $c,c^* $:

\begin{gather}
[c(X),c^* (Y)]=(\bar X, (-\xi_T P_T +\xi_c P_c +\xi_a P_a )K_0 Y) 
\end{gather}
In the smeared form the creation and annihilation operators introduced in this paper are considered as operator valued superdistributions. \\
The commutation relations of the $c,c^* $-operators can be represented in symmetric Fock spaces of supersymmetric functions. The unitarity of the representation depends on the auxiliary constants $\xi $. A proper choice is $\xi_T =\xi_c =\xi_a =1$ in which case the representation is unitary (i.e. it acts in a Hilbert space). Same is true for the choise $\xi_T =1, \xi_c =\xi_a =0$ . \\
The representation of the chiral/antichiral canonical commutation relations (3.1)-(3.4) deserves some comments. For the case $\lambda =0 $ their unitary Fock space representation parallels the representation of the canonical commutation relations used to quantize the non supersymmetric complex scalar field. But for $\lambda $ positive the Fock representation is unusual. It has to be done on tensor products of two component supersymmetric vector functions. The vacuum is the vector function with both components equal to one. In fact it resembles very much the Fock space representation of the canonical commutation relations of the Majorana field (remember that we pass from Dirac to Majorana fermions by identifying pairs of creation and annihilation operators and therefore the Majorana Fock space representations get coupled as compared to the Dirac one).  \\
Let us summarize the results of this section: we provided supersymmetric canonical commutation relations i.e. commutation relations of $a,b,c$ and their conjugates which can be represented in symmetric Fock spaces whose elements are (symmetric) tensor products of supersymmetric functions. The vacuum is the (supersymmetric) function equal to one. One of the main points of this work is the fact that in the chiral/antichiral case (the case of $a, a^* ,b, b^* $) as well as in the properly chosen vector case (for instance the operators $c,c^* $ with commutation relation involving $\xi_T =\xi_c =\xi_a =1 $ or $ \xi_T =1 ,\xi_c =\xi_a =0$) the Fock representations respects positivity i.e. the commutation relations are represented in positive definite Hilbert-Fock spaces (unitary representations). Note further that our representations are supersymmetric invariant because of the invariance of the sesquilinear forms. \\
In the next section we prove that the free supersymmetric quantum fields constructed in a natural way with the help of the supersymmetric creation and annihilation operators are compatible with the standard ones obtained in the supersymmetric literature mainly by path integral methods. In the vector case some particularities appear. But the real interesting aspect, which is not present in the non supersymmetric case, appears in the massless limit especially for the vector field to be studied in section 5.
 
\section{Free Field Supersymmetric Canonical Quantization}

We introduce by means of the creation and annihilation operators  several kinds of supersymmetric free fields (not only the standard ones) which we list below. In the smeared form they are considered as operator valued superdistributions. The definitions to follow are at first sight rather abstract and the natural question to ask is if we get the physically correct supersymmetric fields as result of canonical quantization. The idea behind our construction is to use the two point functions (or commutators) rather than propagators which are at the hart of the functional methods. If the expression of the supersymmetric free field to be defined induces the supersymmetric commutator, then it is declared to be the corresponding quantized supersymmetric field. It turns out that no problems appear in the chiral/antichiral case. In particular the non-propagating auxiliary fields turn out to play no role in the quantization procedure because they are constrained on-shell (in particular they safisfy their algebraic equations of motion) by the strict positivity in our Hilbert space (see Section 2). Problems may appear in the vector case which deserves special attention especially related to the gauge symmetry. These problems in the massless case are studied in Section 5. \\
Let us start with the chiral/antichiral case. First we consider fields generated by the  $a,a^* ,b,b^* $-operators with $\lambda =0 $ :\\
i) The (non-diagonal) chiral field is defined as

\begin{gather}
\varphi (x,\theta ,\bar \theta )=\frac{1}{(2\pi )^{\frac{3}{2}}}\int\frac{d^3 p}{\sqrt 2}[a(p,\theta ,\bar \theta )e^{ipx}+a^* (p,\theta ,\bar \theta )e^{-ipx}]
\end{gather}
ii) The (non-diagonal) antichiral field is defined as

\begin{gather}
\psi (x,\theta ,\bar \theta )=\frac{1}{(2\pi )^{\frac{3}{2}}}\int\frac{d^3 p}{\sqrt 2}[b(p,\theta ,\bar \theta )e^{ipx}+b^* (p,\theta ,\bar \theta )e^{-ipx}]
\end{gather}
The quantum fields act in the Fock spaces of the commutation relations of the $a,a^* $ and $b,b^* $ operators respectively. The antichiral field is the adjoint to the chiral one $\psi =\varphi^* $. In an improper way the operator adjoint $\varphi^* $ can be denoted by $\bar \varphi $ although the bar should be reserved for the conjugate (test) functions. We take permission to use alternatively this improper notation too. We have introduced the fields above only for pedagogical reasons; they seem not to have (in the present massive case) a physical interpretation. \\
Now consider the case $\lambda $ positive (for example $\lambda =1 $ ). In this case we introduce \\
iii) The chiral/antichiral field given by 

\begin{gather}
\phi =
\begin{pmatrix}
\varphi \\
\psi
\end{pmatrix}
\end{gather}
where $\varphi ,\psi =\bar \varphi =\varphi^* $ are defined by the equations above. \\
Finally before going to show that our constructions are physically correct we define the quantized vector field. \\
iv) The vector field is given by

\begin{gather}
V (x,\theta ,\bar \theta )=\frac{1}{(2\pi )^{\frac{3}{2}}}\int\frac{d^3 p}{\sqrt 2}[c(p,\theta ,\bar \theta )e^{ipx}+c^* (p,\theta ,\bar \theta )e^{-ipx}]
\end{gather}
where $c,c^* $ satisfy the canonical commutation relations (3.11).
It depends on the constants $\xi $. Let us fix $\xi_T =\xi_c =\xi_a =1$ or $\xi_T =1, \xi_c =\xi_a =0$. The vector field acts in the Fock space of the $c,c^* $-operators. \\

Now we come to the point of clarifying the relations of the fields introduced above by the method of canonical quantization and those currently used in supersymmetry. We give first the results of the field quantization for the chiral/antichiral in the form of their commutators (or two point functions). We consider 

\begin{gather} \nonumber
\begin{pmatrix} [\bar \varphi_1 ,\varphi_2 ] & [\bar \varphi_1 ,\bar \varphi_2 ] \\ \nonumber
  [\varphi_1 ,\varphi_2 ] & [\varphi_1 ,\bar \varphi_2 ] 
\end{pmatrix}
\end{gather}
computed with the help of the Fock representation of their canonical commutation relations. Here $\varphi (z_i )=\varphi_i ,\bar \varphi (z_i )=\bar \varphi_i ,i=1,2 $. Leaving out the usual commutator $\Delta $- factor it follows from (4.1)-(4.3) that they are  

\begin{gather}
\begin{pmatrix} \frac{1}{16}D^2 \bar D^2 & \frac{m}{4}D^2 \\
  \frac{m}{4}\bar D^2 & \frac{1}{16}\bar D^2 D^2 
\end{pmatrix} \delta^2 (\theta_1 -\bar \theta_1 )\delta^2 (\theta_2 -\bar \theta_2 )
\end{gather}
In the massive case using scaled canonical commutation relations for $a,a^* ,b,b^* $ (i.e. scaling (3.1)-(3.4) by the inverse d'Alembertian) we can arrange to obtain instead (4.5)

\begin{gather}
\begin{pmatrix}  P_c & P_+ \\
                 P_- & P_a 
\end{pmatrix} \delta^2 (\theta_1 -\bar \theta_1 )\delta^2 (\theta_2 -\bar \theta_2 )
\end{gather}
For the vector case we get (in the case $\xi_c =\xi_a =\xi_T =1 $) up to the $\Delta $- factor the commutator $[V_1 ,V_2 ]$ as:

\begin{gather}
(-P_T +P_c +P_a  )\delta (\theta_1 -\bar \theta_1 )\delta (\theta_2 -\bar \theta_2 )=(1-2P_T )\delta (\theta_1 -\bar \theta_1 )\delta (\theta_2 -\bar \theta_2 )
\end{gather}
The result for the chiral/antichiral field shows that we have obtained, as expected, the results obtained at the level of propagators by path integral methods or by brute force using the propagators (or the two point functions) of the multiplet components \cite{WB} p.72. Consequently it is possible to expand $ a(p,\theta ,\bar \theta ), a^* (p,\theta ,\bar \theta ), b(p,\theta ,\bar \theta ), b^* (p,\theta ,\bar \theta ) $ in $\theta ,\bar \theta $-variables with coefficients being the usual (Lorenz covariant) creation and annihilation operators of the multiplet components.
For the case of the massive vector field the result (4.7) coincide with the path-integral result \cite{WB} p.73 but there are some discrepances with the result computed on component multiplets. To see this explicitly we must compute in (4.7) $-P_T +P_c +P_a =1-2P_T $ applied to the delta funtion $\delta^2 (\theta_1 -\theta_2 )\delta^2 (\bar \theta_1 -\bar \theta_2 )$. We prefer not to use the exponential form of $P_T$ \cite{WB} but use directly explicit formulas for the projector $P_T $ applied to the first variables $\theta_1 ,\bar \theta_1 $ of $\delta^2 (\theta_1 -\theta_2 )\delta^2 (\bar \theta_1 -\bar \theta_2 ) $. We expand first in $\theta_1 ,\bar \theta_1 $ as follows

\begin{gather}\nonumber
\delta^2 (\theta_1 -\theta_2 )\delta^2 (\bar \theta_1 -\bar \theta_2 )=\theta_2 ^2 \bar \theta_2 ^2+\theta_1 (-2\theta_2 \bar \theta_2 ^2 )+ \\ \nonumber
+\bar \theta_1 (-2\bar \theta_2 \theta_2 ^2 )+ \theta_1 ^2 \bar \theta_2 ^2 +\bar \theta_1 ^2 \theta_2 ^2 +2(\theta_1 \sigma^l \bar \theta_1 )(\bar \theta_2 \bar \sigma_l \theta_2 )+\\
+\theta_1^2 \bar \theta_1 (-2\bar \theta_2 )+\bar \theta_1^ 2 \theta_1 (-2\theta_2 )+\theta _1^2 \bar \theta_1 ^2 
\end{gather}
Here we have used the simple identity

\begin{gather}
2(\theta_1 \theta_2 )(\bar \theta_1 \bar \theta_2 )=(\theta_1 \sigma^l \bar \theta_1 )(\bar \theta_2 \bar \sigma_l \theta_2) 
\end{gather}
Now we apply formula (2.30) from \cite{C1} to get 

\begin{gather} \nonumber
(1-2P_T )\delta^2 (\theta_1 -\theta_2 )\delta^2 (\bar \theta_1 -\bar \theta_2 )\Delta = \\ \nonumber
=[\frac{4}{\square }(1-i\theta_1 \sigma^l \bar \theta_2 \partial_l -i\bar \theta_1 \bar \sigma^l \theta_2 \partial_l ) + \\ \nonumber
+\theta_1 ^2 \bar \theta_2 ^2 +\bar \theta_1 ^2 \theta_2 ^2 +2(\theta _1 \sigma^l \bar \theta_1 )(\theta _2 \sigma_l \bar \theta_2 +\frac{2}{\square }\partial_l \partial^m \bar \theta_2 \bar \sigma_m \theta_2 )- \\
-i\theta_1 ^2 \bar \theta_1 \bar \sigma^l\theta_2 \bar \theta_2 ^2 \partial_l -i\bar \theta_1 ^2 \theta_1 \sigma^l \bar \theta_2 \theta_2 ^2 \partial_l +\frac{1}{4}\square \theta_1 ^2 \bar \theta_1 ^2 \theta_2 ^2 \bar \theta_2 ^2 ]\Delta 
\end{gather}
On the other hand we write down the most general vector field as
\begin{gather}\nonumber 
V(z)=V(x,\theta ,\bar \theta )= \\ \nonumber
=f(x)+\theta \varphi (x) +\bar \theta \bar \chi (x)+\theta ^2 m(x)+\bar \theta^2 n(x)+ \\ 
+\theta \sigma^l \bar \theta v_l (x)+\theta^2 \bar \theta \bar \lambda (x)+\bar \theta^2 \theta \psi (x)+ \theta^2 \bar \theta^2 d(x)
\end{gather}
The coefficients of the Grassmann variables in $V(z)$ are quantum fields. This is the difference between $V(z)$ and the supersymmetric test function $X(z)$. For a real vector field we must impose the conditions
 \[ f=\bar f,\chi =\varphi ,n=\bar m, \lambda =\psi ,d=\bar d \]
Performing a standard elementary computation we can determine the commutator $[V_1 ,V_2 ]=[V(z_1 ),V(z_2 )]=[V(x_1 ,\theta_1 ,\bar \theta _1 ),V(x_2 ,\theta_2 ,\bar \theta _2 )]$
of $V(z)$ using the commutators of the components. By using standard commutation relations for the multiplet components the result is (the usual commutator is abreviated with $\Delta $):

\begin{gather}\nonumber
[V_1 ,V_2 ]=[1 
-i\theta_1 \sigma^l \bar \theta_2 \partial_l -i\bar \theta_1 \bar \sigma^l \bar \theta_2 \partial_l + \\ \nonumber
+\theta_1 ^2 \bar \theta_2 ^2 +\bar \theta_1 ^2 \theta_2^2 +(\theta_1 \sigma^l \bar \theta_1 )(\theta_2 \sigma^m \bar \theta_2 )(\eta_{lm}-\frac{1}{m^2 }\partial_l \partial_m ) - \\ 
-i\theta_1 ^2 \bar \theta_2 ^2 \bar \theta_1 \bar \sigma^l \theta _2 \partial_l - i\bar \theta_1 ^2 \theta_2 ^2 \theta_1 \sigma^l \bar \theta _2 \partial_l +\theta_1 ^2 \bar \theta_1 ^2 \theta_2 ^2 \bar \theta_2 ^2 ]\Delta
\end{gather}
We get consistency but not coincidence with the result (4.10) which in turn, in our setting up, was induced from the canonical commutation relations (3.11) with  $\xi_T =\xi_c =\xi_a =1$. Besides different normalization factors it turns out that the Majorana fermions must be non-diagonal. They are defined in (4.14). If we would try to use in $V(z)$ usual Majorana fermions for $\varphi ,\bar \varphi $ and $\lambda ,\bar \lambda $ defined in (4.13) instead of non-diagonal ones (4.14) then in the massive case (4.12) would acquire unwanted extra terms and it would be impossible to assure the expected positivity i.e. the positivity induced by the representations of the supersymmetric canonical commutation relations. We provide the necessary details. Usual Majorana fermions $\varphi ,\bar \varphi $ have commutators given by

\begin{gather}\nonumber
[\varphi_{\alpha }(x),\varphi^{\beta }(x')]=i\delta_{\alpha }^{\beta }m\Delta(x-x') \\ \nonumber
[\bar \varphi^{\dot \alpha }(x),\bar \varphi_{\dot \beta }(x')]=i\delta^{\dot \alpha }_{\dot \beta }m\Delta(x-x') \\ 
[\varphi_{\alpha }(x),\bar \varphi_{\dot \beta }(x')]=\sigma_{\alpha \dot \beta }^l \partial_l \Delta(x-x') 
\end{gather}
whereas our non-diagonal (massive or massless) Majorana fermions are defined by

\begin{gather}
[\varphi_{\alpha }(x),\varphi^{\beta }(x')]=[\bar \varphi^{\dot \alpha }(x),\bar \varphi_{\dot \beta }(x')]=0  
\end{gather}
with the $[\varphi_{\alpha },\varphi^{\beta }]$ commutator as above.
The commutator $[V_1,V_2]$ in (4.12) was computed using (4.14). Unwanted contributions in (4.12) appear if we insist on (4.13) with non-vanishing diagonal commutators $[\varphi_{\alpha },\varphi^{\beta }],[\bar \varphi^{\dot \alpha },\bar \varphi_{\dot \beta }]$. \\
As far as different normalizations are concerned, they are harmless because they can be transfered to the corresponding canonical commutation relations for the component fields or absorbed in field redefinitions (without practical gain). Such factors have been reported in the supersymmetric literature (see for example \cite{WB} p.73). There is still a point to remark: even if we work with non-diagonal Majorana fermions the combination 
\[2(\theta _1 \sigma^l \bar \theta_1 )(\theta _2 \sigma_l \bar \theta_2 +\frac{2}{\square }\partial_l \partial^m \bar \theta_2 \bar \sigma_m \theta_2 )=(\theta_1 \sigma^l \bar \theta_1 )(\theta_2 \sigma^m \bar \theta_2 )(2\eta_{lm}-\frac{4}{\square }\partial_l \partial_m ) \] 
from (4.10) is not quite exactly reproduced in (4.12). In the zero mass limit the discrepances tend to vanish (for example the massless Majorana fermions are non diagonal) but we cannot make a definite statement in this section because as expected the massless limit of the vector field doesn't exist by brute force. Some degree of sophistication is required to be explained in the next section.\\       
Concluding our disscusion we see that facing the supersymmetric canonical quantization with computations based on multiplet expansion there are some particularities in the case of the supersymmetric massive vector field (use of non-diagonal Majorana fermions, several normalization factors and a discrepance in the commutator of the non-supersymmetric vector field). As a consequence expansions of $ c(p_1 ,\theta_1 ,\bar \theta_1 ), c^* (p_2 ,\theta_2  ,\bar \theta_2 ) $ in the $\theta ,\bar \theta $-variables do not reproduce creation and annihilation operators of the multiplet components and are rather obscure. More interesting aspects of the supersymmetric canonical quantization appear in the case of the massless vector field to which we turn now.

\section{The massless case} 

Now we study the canonical commutation relations and the fields generated by them in the massless case under the special condition of gauge symmetry. It is well known that the limit of vanishing mass for scalar and Dirac quantum free field in four dimensional space-time is well defined. On the contrary in the vector field case this limit is not well defined. In order to pass to zero mass maintainig positivity, Lorentz invariance (and if necessary) gauge invariance we need some amendments which we can impose at the level of fields (like for instance Gupta-Bleuler, Nielsen-Lautrup or even the Kugo-Ojima procedure over ghosts). An equivalent possibility which is less known uses amendments at the level of the test functions to which the fields are to be applied \cite{SW}. We prefer in this paper the second procedure not only for reasons of taste. Certainly by the main property of distribution theory the two procedures are equivalent (see for instance the short disscussion at the end of section 2 in \cite{C1}). \\
We give the arguments for the case of the vector field. At the first sight it appears that we are in trouble because the d'Alembertian blows up in the projection denominators. But there is an elegant way out of this dilemma which remembers the "subsidiary" condition of the Gupta-Bleuler quantization applied at the level of test functions. The physical (positive definite) Hilbert space is produced by a restriction followed by a factorization (and completion) of the general space of regular (for instance in $S$) test superfunctions. We provide the necessary details. Let us first restrict the space of test function to the linear subset characterized by the relations \cite{C1}
 
\begin{gather}
d(x)=\square D(x) \\
\bar \lambda (x)=\bar \sigma^l \partial_l \Lambda (x) \\ 
\psi (x) =\sigma^l \partial_l \bar \Psi (x) \\
v^l (x)=\partial ^l \rho (x)+\omega^l (x) 
\end{gather}
where $ D(x),\Lambda (x),\Psi (x),\rho (x),\omega (x) $
are arbitrary functions of some regularity  and $\partial_l \omega^l (x)=0 $.
Because of lack of better terminology we named the corresponding functions special supersymmetric. 
At the first sight it appears that (5.1)-(5.4) do not restrict the component functions. Indeed (5.1)-(5.4) can be looked at as differential equations for the unknown funtions $D,\Lambda ,\bar \Psi ,\rho ,\omega $. But if we require solutions with some regularity (for example elements in the Schwatz function space) then it follows that (5.1)-(5.4) are true restrictions. 
Using the formulas (2.10), (2.28)-(2.30) from \cite{C1,C2} it is easy to see that the projections $P_i ,i=c,a,T $ transform the linear space of special supersymmetric functions into itself. The same formulas show that restricting to the special supersymmetric functions we cancel the blowing up d'Alembertian in the projection denominators. Now there is a sector overlap i.e. the projection operators $P_i $ are not disjoint. The sector overlap was computed in section 3 of \cite{C1}. For our purposes it is not necessary to know how it looks like. We factorize the sector overlap (chiral, antichiral and transversal) and after completion obtain a space of supersymmetric test functions whith unchanged scalar product which can accomodate the sector projections as bona fide disjoint projection operators. They act in the space of special supersymmetric functions. The quantized massless vector field is now defined by its two point function or commutator as the massless limit in physical Hilbert space (!) of the two point function or commutator (4.7) and (4.10) of the massive vector field considered on special supersymmetric functions. The last one was defined via unitarily represented canonical commutation relations in Section 4. We stress that the blowing up d'Alembertians in (4.10) are absorbed in the process of integration against special supersymmetric functions (the Hilbert space limit implies this integration). \\
Let us show explicitely that in the process of quantization we really obtain the same result given by the functional integral methods. First note that our construction respects gauge symmetry. Indeed arbitrary constants $\xi_i $ in (3.13) are related to several gauges for the vector field. The well-known Wess-Zumino gauge corresponds to $\xi_c =\xi_a =0, \xi_T =1$ because it cancells the chiral/antichiral part of the vector field commutator (see (4.7) in which $P_c ,P_a $ disappear). In physical terms defining the (real) vector field in the Wess-Zumino gauge we put forward the (formal) field \cite{WB}

\begin{gather}
V(z)=-\theta \sigma^l\bar \theta v_l(x)+i\theta^2\bar \theta \bar \lambda(x)-i\bar \theta^2\theta \lambda (x)+ \frac{1}{2} \theta^2 \bar \theta^2 D(x)
\end{gather}
with the divergence condition on $v_l $ and not the transversal sector on which $P_T $ projects. By the very nature of distribution theory test functions in the transversal sector of the (strict positive definite) Hilbert space are equivalent with the expression (5.5) of the physical vector field proving the equivalence of our construction with the usual one. Certainly arbitrary $\xi_i $ provide gauge symmetry of the quantization inside the family of gauges induced by the two-point function

\begin{gather}
(-\xi_T P_T +\xi_c P_c +\xi_a P_a  )K_0 (z_1 -z_2 )
\end{gather}
(for $K_0 (z_1 -z_2 ) $ see (2.8))\\
For the convenience of the reader we remark that the present situation is similar to the quantization of the non-supersymmetric massless vector field. In this case we have, instead of three, only two projections; $P_L $ (longitudinal photons) and $P_T $ (transversal and scalar photons) acting between vector test functions ($P_L= P_{lm }=\frac{\partial_l\partial_m}{\square }$). It is nice that they sum up to one, $P_L +P_T =1$, but the price to pay is indefiniteness. Moreover they are responsible for the celebrated Krein (indefinite) structure of electrodynamics. All unpleasant aspects besides indefiniteness (d'Alembertians in the denominators and sector overlap which corresponds to the gradient gauge ambiguity and has to be factorized) are already present. From the technical point of view the Lorentz divergence condition $\partial_l v^l =0$ allows us to pass from the indefinite to the semidefinite, and the factorization of the gradient from the semidefinite to the stricty definite metric. There are two well-known (Lorentz invariant) possibilities to restore positivity killing at the same time the unpleasant appearances: either by using the divergence condition followed by the gradient factorization (which cancels the longitudinal sector) or by using the kernel induced by the difference $P_L -P_T $ applied to the scalar two point function (or commutator). The first possibility is the rigorous variant of the Gupta-Bleuler quantization \cite{SW}. At least by now it should be clear that our restrictions (5.1)-(5.4) are the supersymmetric couterpart of the divergence condition in the quantization process of the vector field.\\
We come to the end of this section and therefore it is time to ask ourselves if the situation described regarding the massless vector field is, at least in principle, different or not from the usual case in non-supersymmetric quantum field theory. By a closer look we realize that there is not too much difference. The new point is the restriction (5.1)-(5.4) to the special supersymmetric functions. For ( $\xi_c =\xi_a =0 $) it is equivalent to the Wess-Zumino gauge and Lorentz divergence condition which in the usual case eliminates the longitudinal modes of the photon. On the other hand the restriction combined with the choice $ \xi_c =\xi_a =\xi_T =1 $ corresponds to the Stueckelberg quantization in the unitary gauge. The non-unitary Feynman gauge would correspond to $ \xi_c =\xi_a =1, \xi_T =-1 $ and no restriction on supersymmetric functions (the projections summ up to one and there is no reason to deal with the inverse d'Alembertian). From a technical point of view the canonical quantization of the supersymmetric massless vector field is less trivial (but by no means more complicated) than the well-known Gupta-Bleuler or Stueckelberg procedures.

\section{Comments} 

We have introduced unitarily represented supersymmetric canonical commutation relations and have recovered by canonical quantization the free supersymmetric massive and massless chiral/antichiral and vector fields. The tool of our approach was the supersymmetric Hilbert-Krein structure \cite{C1}. The procedure parallels very closely the usual one in non-supersymmetric quantum field theory. Non-propagating auxiliary fields are harmless. The point of the paper is to prove that canonical quantization of massive and massless free supersymmetric fields by means of unitarily represented supersymmetric canonical commutation relations is possible in a natural way. This includes in the massless case a non-trivial restriction to a subspace of supersymmetric test functions. \\
Before ending let us remark that in the non-supersymmetric case there is ample literature dedicated to the canonical formalism of gauge theories related to the BRST formalism which via ghosts and antighosts manage to include unitarity (positivity) at a certain level of rigorousity (in particular the Kugo-Ojima theory). The natural question is if such achievements can be taken over to the supersymmetric case. The present paper can be regarded as a preamble to such developements. Whereas the general case of gauge theories has to be still carefully analized, the situation is relatively simple if the ghosts decouple from fields as this turns out to happen in electrodynamics. In this case the ghosts are free and the BRST charge $Q$ can be easily constructed with the help of the creation and annihilation operators (see for instance \cite{W}, vol 2, p.33-34). Only the kernel $ker Q$ of the operator $Q$ is responsible for the physical states of the theory. The procedure can be easily taken over to the supersymmetric case. Using the results of Section 4 of this paper and changing the commutators in (3.1)-(3.6) into anticommutators for the non-diagonal setting up ($\lambda =0$) we construct the supersymmetric BRST charge $Q$. The corresponding supersymmetric ghost Fock space is antisymmetric. The quantum ghosts are not causal. The ghost quantization of the massless vector field (and with some modifications of the massive too) is unitary equivalent to the quantization disscussed in Section 5. The rigorous proof of this statement is even simpler in the supersymmetric than in the non-supersymmetric case (for the non-supersymmetric case see the review \cite{G}).

\end{document}